\documentclass[preprint,12pt]{elsarticle}

\usepackage{amssymb}
\usepackage{float}
\usepackage{subcaption}
\usepackage{multirow}
\usepackage{amsmath}
\usepackage{array}

\begin{document}

\begin{frontmatter}



\title{Propellant Discovery For Electrospray Thrusters Using Machine Learning}


\author[inst1]{Rafid Bendimerad}
\author[inst1]{Elaine Petro}

\affiliation[inst1]{organization={Cornell Sibley School of Mechanical and Aerospace Engineering},
            addressline={124 Hoy Rd}, 
            city={Ithaca},
            postcode={14850}, 
            state={NY},
            country={United States of America}}

\begin{abstract}
This study introduces a machine learning framework to predict the suitability of ionic liquids with unknown physical properties as propellants for electrospray thrusters based on their molecular structure. We construct a training dataset by labeling ionic liquids as suitable (+1) or unsuitable (-1) for electrospray thrusters based on their density, viscosity, and surface tension. The ionic liquids are represented by their molecular descriptors calculated using the Mordred package. We evaluate four machine learning algorithms—Logistic Regression, Support Vector Machine (SVM), Random Forest, and Extreme Gradient Boosting (XGBoost)—with SVM demonstrating superior predictive performance. The SVM predicts 193 candidate propellants from a dataset of ionic liquids with unknown physical properties. Further, we employ Shapley Additive Explanations (SHAP) to assess and rank the impact of individual molecular descriptors on model decisions.
\end{abstract}

\begin{graphicalabstract}
\includegraphics[width=\linewidth]{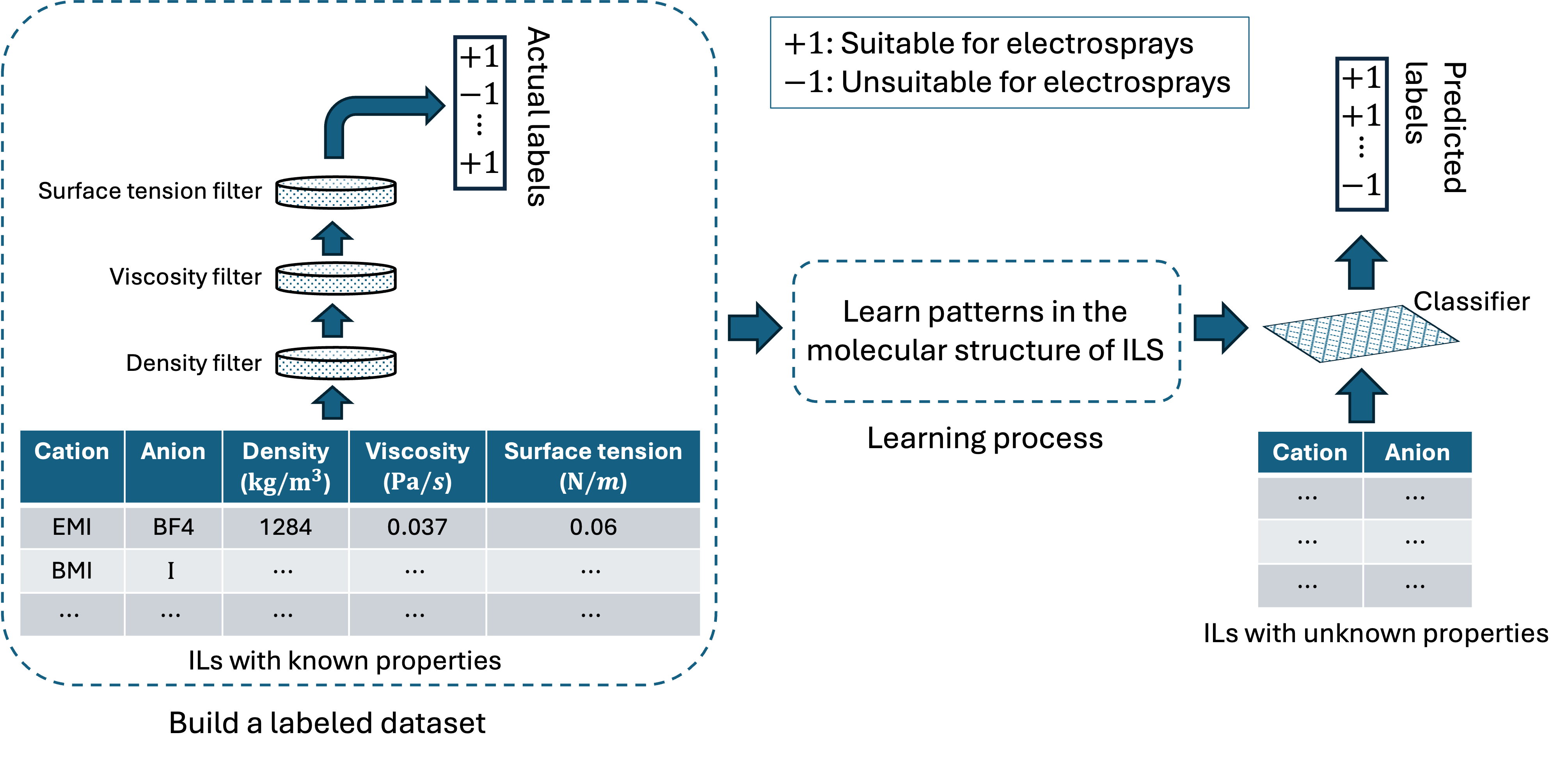}
\end{graphicalabstract}

\begin{highlights}

\item This paper introduces a general machine learning framework designed to predict the suitability of ionic liquids with unknown physical properties for specialized applications in engineering.

\item The utility of this framework is demonstrated for an application in aerospace engineering where ionic liquids with specific physical properties are required to serve as propellants for electrospray thrusters.

\item This framework produces a classifier that predicts 193 candidate ionic liquids that could potentially be used as propellants for electrospray thrusters. 

\end{highlights}


\end{frontmatter}


\section{Introduction} \label{sec:introduction}

Ionic liquids are organic salts composed of cations and anions that exist in a liquid state at room temperatures \cite{freemantle2010introduction}. Due to their distinctive physicochemical properties, such as low volatility, high thermal stability, and wide electrochemical windows, they garner significant attention in various scientific and industrial applications such as solvents, electrolytes, lubricants, catalysts, drug delivery systems, absorption chillers, and many other applications   \cite{greer2020industrial,welton2018ionic,hallett2011room,zhou2009ionic,shamshina2013ionic,koutsoukos2021review}. Despite their versatility, the applicability of ionic liquids is contingent upon the fulfillment of specific physicochemical criteria required by each application. Therefore, the precise selection of an ionic liquid tailored to meet the demands of a particular application is imperative for ensuring effective and efficient performance.

However, identifying all the suitable ionic liquid candidates for a specific application is not a simple task as  hundreds of ionic liquids are already available commercially, with potentially millions more awaiting synthesis. Given the current advancements in organic synthesis, which present virtually no boundaries, a myriad of ionic liquids can be produced \cite{paduszynski2019extensive}. A comprehensive experimental study examining the effects of the chemical structure of ILs on their pertinent properties is not practically achievable. Therefore, it is essential to exploit the sparse experimental data that are available to extrapolate properties of ILs that have yet to be empirically characterized.

Former studies have utilized the group contribution method (GCM) within a quantitative structure-property relationship (QSPR) framework for estimating density \cite{paduszynski2019extensive}, viscosity \cite{paduszynski2019extensive2}, and surface tension \cite{paduszynski2021extensive} of ionic liquids. The group contribution method involves assigning numerical values to specific functional groups within a molecule based on their known contribution to certain properties. The property of a compound is estimated as a summation of the  contributions of simple first-order groups \cite{constantinou1994new}. However, the first step of the GCM involves identifying specific functional groups that are known to have influence on the property of interest, which requires prior knowledge and effort. Furthermore, GCM focuses on the contribution of groups rather than the detailed molecular structure. These limitations can be overcome by the utilization of molecular descriptors calculated from the SMILES (Simplified Molecular Input Line Entry System) representation of the molecules. SMILES is a chemical notation system that compactly represents molecular structures as linear ASCII strings, encapsulating atomic constituents, bond configurations, and stereochemistry in a computationally interpretable format \cite{weininger1988smiles}. The SMILES format allows for representing molecules in a molecular graph, which in turn can be used for the calculation of a comprehensive array of physical descriptors that capture detailed structural information.  The utilization of molecular descriptors eliminates the need to select suitable functional groups and enables capturing a broader range of properties, including geometric, electronic, and topological characteristics, crucial for detailed molecular analysis.

Machine learning approaches based on the SMILES representation have been used for the study of organic molecules and polymers. Pinheiro et al.~used a feed-forward neural network (FNN) to predict nine molecular properties of organic molecules based on their SMILES representation \cite{pinheiro2020machine}. Chen et al.~utilized a chemical language processing model to predict polymers’ glass transition temperature using a recurrent neural network (RNN) that receives the SMILES strings of a polymer’s repeat units as inputs \cite{chen2021predicting}. On the other hand, Saini used molecular descriptors calculated from SMILES to predict the empirical polarity of organic solvents using an artificial neural network \cite{saini2023machine}.

Inspired by these applications to organic molecules and polymers, and driven by the need to identify new ionic liquid candidates for industrial and engineering applications, we propose a machine learning framework to classify ionic liquids as suitable or unsuitable for a given application. To demonstrate the utility of our approach, we apply our framework to solve a problem in aerospace engineering. Specifically, we aim to discover new ionic liquids with specific physical properties that can serve as propellants for electrospray thrusters.

\begin{figure}[ht!]
    \centering
    \includegraphics[width=0.7\linewidth]{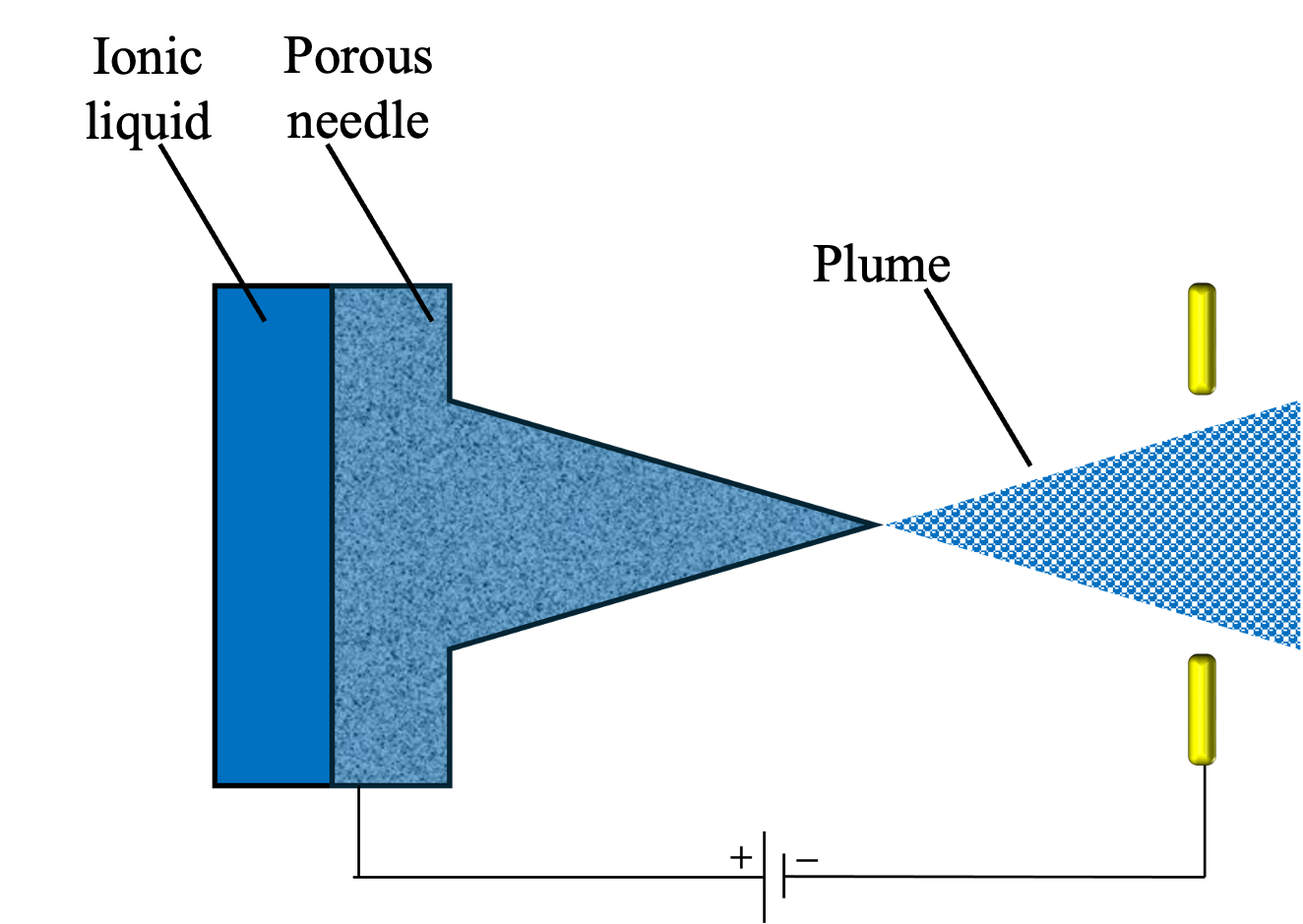}
    \caption{Diagram of an electrospray thruster consisting of a porous needle containing an ionic liquid and a downstream extractor plate.}
    \label{fig:electrospray}
\end{figure}

Electrospray thrusters have emerged as a pivotal technology for micropropulsion of small satellites. They are particularly attractive because of their compactness, simplicity, scalability, and high specific impulse \cite{Krejci2015design}. These thrusters employ very strong electric fields to eject ions from ionic liquids at high velocity, generating thrust \cite{lozano2005ionic}.  A diagram of an electrospray thruster with a porous needle containing an ionic liquid is shown in Fig.~\ref{fig:electrospray}. For the optimal selection of an ionic liquid in electrospray thrusters, it is imperative to consider both the attributes of the bulk liquid and the characteristics of the cation and anion. Essential properties of the bulk liquid encompass density, surface tension, electrical conductivity, viscosity, and electrical permittivity. On the other hand, the key ion attributes are its molecular complexity, structure, and charge distribution \cite{miller2019characterization}. The final objective of this paper is to provide a list of candidate ionic liquids that have the potential to serve as propellants for electrospray thrusters.


\section{Materials and Methods} \label{sec:methods}

\begin{figure}[ht!]
    \centering
    \includegraphics[width=0.9\linewidth]{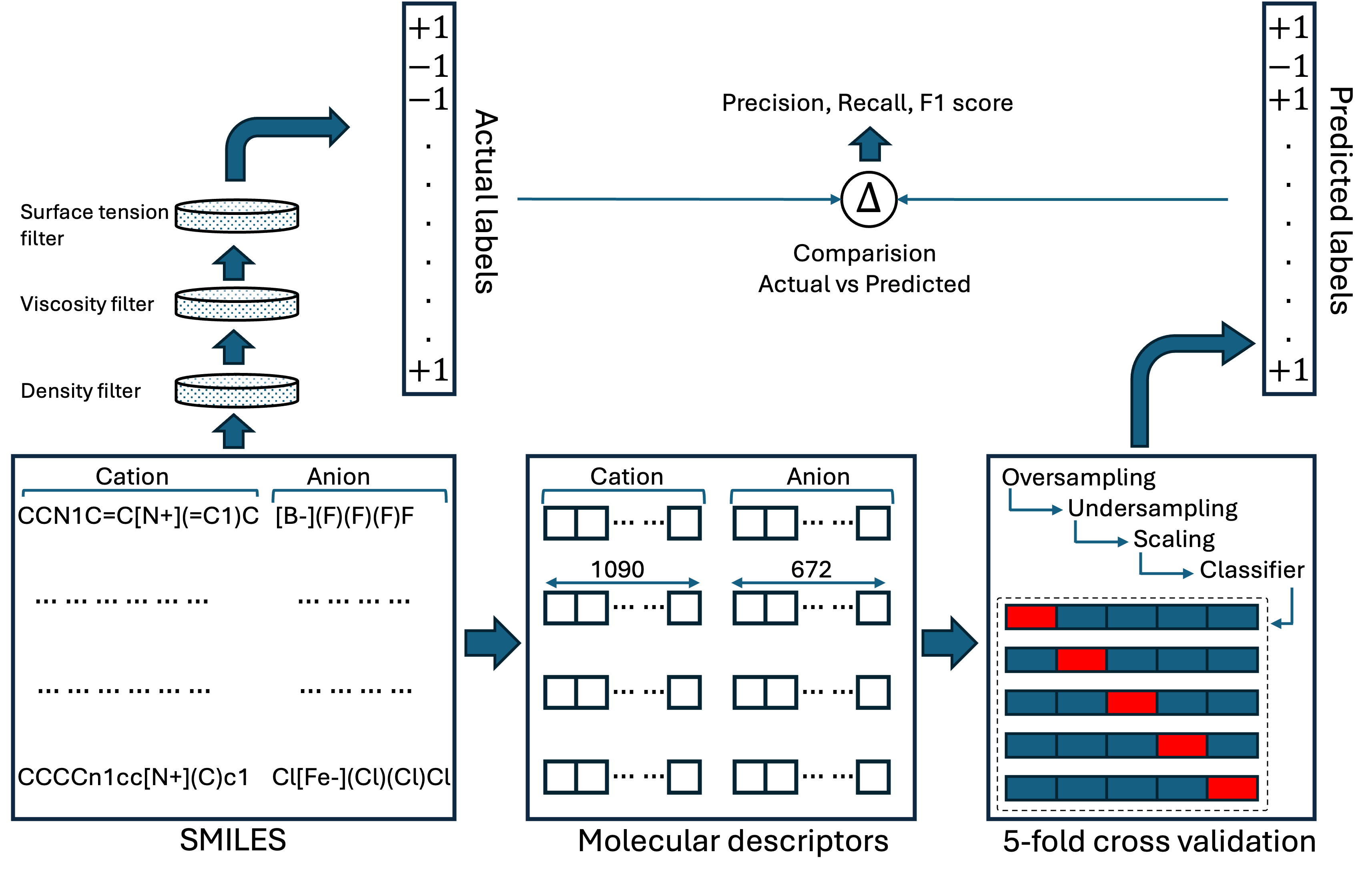}
    \caption{Diagram of the model architecture: from the SMILES representation to label prediction and performance evaluation.}
    \label{fig:workflow}
\end{figure}

We will label ionic liquids as suitable (+1) or unsuitable (-1) based on their density, viscosity, and surface tension. Molecular descriptors for each ionic liquid will be computed from their SMILES representation using the Mordred library \cite{moriwaki2018mordred}. These descriptors, along with the labels, will form our training dataset. We will train various classifiers and evaluate their performance; the most effective classifier will then be applied to a larger dataset with unknown physical properties to identify potential new propellants for electrospray thrusters. The workflow of this approach is shown in Fig.~\ref{fig:workflow}.

\subsection{Database}

To ensure stable emission at high currents, the propellant for electrospray thrusters must meet specific physicochemical criteria. These include an electrical conductivity above 1 S/m, a Gibbs free energy less than 1.2 eV, an electrochemical window exceeding 2.5 V, and a surface tension above 0.05 N/m. Furthermore, the density should be as high as possible whereas the viscosity should be as low as possible \cite{miller2019characterization, gallud2023studies}.

We set the lower limit of the density to 1284 kg/$\text{m}^3$, which is the density at room temperature of EMIBF$_4$ that is the most common ionic liquid used for electrospray propulsion. On the other hand, we set the higher limit of viscosity to 0.060 Pa$\cdot$s, which corresponds to the viscosity at room temperature of EMIFAP that has already been fired in previous experiments \cite{miller2019characterization}. We intentionally set a relatively low lower boundary for density and a relatively high upper boundary for viscosity in order to cover a wider range of acceptable ionic liquids from our dataset.

Because of the unavailability of extensive datasets on electrical conductivity, Gibbs free energy, and electrochemical window, this study only accounts for the density, the viscosity, and the surface tension. For this, we will use the three datasets provided by Paduszynski \cite{paduszynski2019extensive, paduszynski2019extensive2, paduszynski2021extensive}. The first dataset includes the densities of 1161 ionic liquids, the second includes the viscosities of 1974 ionic liquids, and the third includes the surface tensions of 542 ionic liquids. Collectively, these datasets account for 2261 unique ionic liquids. From this aggregated dataset, only 14 ionic liquids meet all specified criteria for density, viscosity, and surface tension and will be labeled as positive examples (+1). Conversely, 1782 ionic liquids fail to meet at least one criterion and will be labeled as negative examples (-1). The remaining 465 ionic liquids meet one or two of the criteria; however, the data for the third property is incomplete, and therefore the ionic liquid cannot be labeled. The 14 ionic liquids with a positive label are reported in Table~\ref{tab:14pos}. It should be noted that the names of the cations in this table  explicitly include the presence of hydrogen atoms on the nitrogen, such as `1-ethyl-3-methyl-1H-imidazol-3-ium' instead of the more commonly used `1-ethyl-3-methylimidazolium'.

\begin{figure}[ht!]
    \centering
    \includegraphics[width=\linewidth]{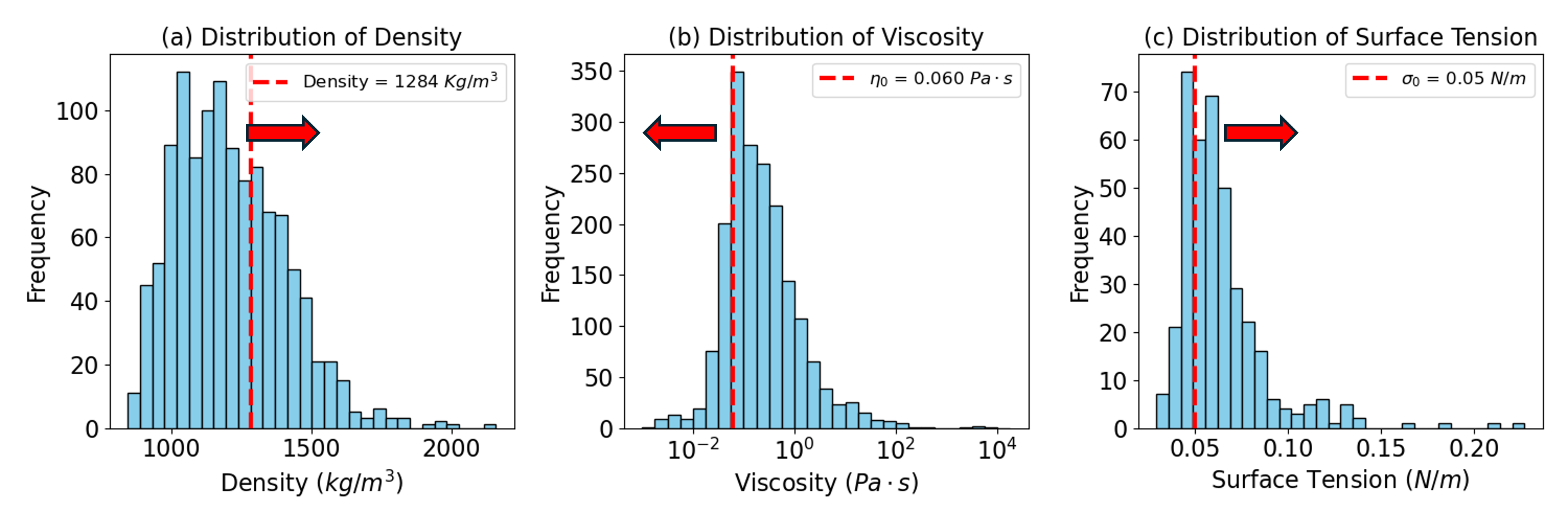}
    \caption{Histogram of (a) the density, (b) the viscosity, and (c) the surface tension in the dataset. The red dashed line represents the required threshold.}
    \label{fig:data}
\end{figure}

\begin{figure}[ht!]
    \centering
    \includegraphics[width=1\linewidth]{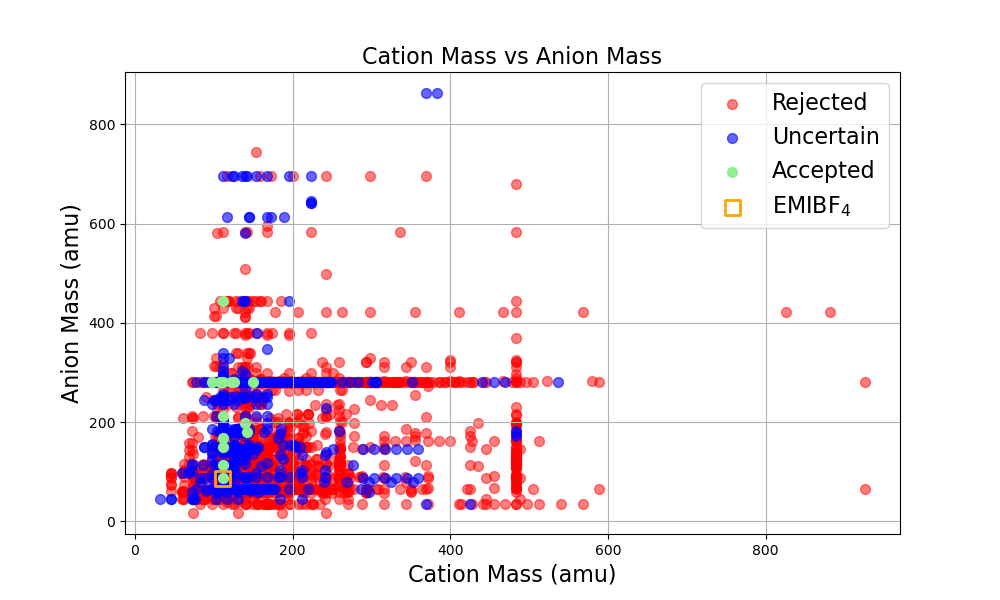}
    \caption{Mass of the anion the cation of the accepted, rejected, and uncertain ionic liquids.}
    \label{fig:accepted_rejected}
\end{figure}

The distributions of the density, viscosity, and surface tension of the ionic liquids in the dataset are shown in Fig.~\ref{fig:data}. The masses of the cation/anion pairs of the ionic liquid are shown in Fig.~\ref{fig:accepted_rejected}. The figure shows the 14 ionic liquids that satisfy all the requirements in green (accepted), the 1782 ionic liquids that do not satisfy at least one requirement in red (rejected), and the 465 ionic liquids that could not be classified because of missing data about one of their physical properties in blue (uncertain). The figure additionally highlights EMIBF$_4$ within a square, which is the most frequently utilized ionic liquid for electrospray thrusters.

\begin{table}[ht!]
\renewcommand{\arraystretch}{1.2}
\centering
\caption{List of ionic liquids that satisfy the physical requirements.}
\label{tab:additional_ions}
\begin{tabular}{|l|l|}
\hline
\textbf{Cation} & \textbf{Anion} \\
\hline
1,3-dimethyl-1H-imidazol-3-ium & bis(trifluoromethylsulfonyl)azanide \\ \hline
1-butyl-1-methylpyrrolidin-1-ium & bis(flurosulfonyl)azanide \\ \hline
1-butyl-3-methyl-1H-imidazol-3-ium & tetrachloroferrate (III) \\ \hline
1-butyl-4-methylpyrrolidin-1-ium & bis(trifluoromethylsulfonyl)azanide \\ \hline
1-ethyl-3-methyl-1H-imidazol-3-ium & bis(trifluoromethylsulfonyl)azanide \\ \hline
1-ethyl-3-methyl-1H-imidazol-3-ium & tetrachloroaluminate \\ \hline
1-ethyl-3-methyl-1H-imidazol-3-ium & tetrachlorogallate \\ \hline
1-ethyl-3-methyl-1H-imidazol-3-ium & tetrafluoroborate \\ \hline
1-ethyl-3-methyl-1H-imidazol-3-ium & trifluoroacetate \\ \hline
1-ethyl-3-methyl-1H-imidazol-3-ium & trifluoromethanesulfonate \\ \hline
1-ethyl-3-methyl-1H-imidazol-3-ium & trifluorotris(pentafluoroethyl)-$\lambda^5$-phosphanuide \\ \hline
1-ethylpyridin-1-ium & bis(trifluoromethylsulfonyl)azanide \\ \hline
1-methyl-3-propyl-1H-imidazol-3-ium & bis(trifluoromethylsulfonyl)azanide \\ \hline
1-propylpyridin-1-ium & bis(trifluoromethylsulfonyl)azanide \\
\hline
\end{tabular}
\label{tab:14pos}
\end{table}
 
\subsection{Calculation of molecular descriptors}

In this subsection, we detail the process used for calculating molecular descriptors from the SMILES representations of cations and anions employing the RDKit \cite{landrum2006rdkit} and Mordred \cite{moriwaki2018mordred} packages. The initial step involves converting SMILES strings into molecular structures using RDKit. This conversion results in a graph-based representation where atoms are represented as vertices and bonds are represented as edges, effectively capturing the molecular topology. Then, the Mordred package calculates the molecular descriptors of the cations and the anions from their molecular structure. The Mordred package is a comprehensive cheminformatics tool designed to compute a wide array of molecular descriptors. It can generate both two-dimensional (2D) and three-dimensional (3D) descriptors that capture various molecular properties essential for our analysis \cite{moriwaki2018mordred}.   An example of a 2D descriptor is the topological radius, which represents the maximum graph-theoretical distance from any atom to the centroid of the molecule. On the other hand, an example of a 3D descriptor is the geometric radius, which represents the actual physical distance from any atom to the geometric center of the molecule in three-dimensional space.

For each cation and anion, we computed a total of 1613 2D descriptors and 213 3D descriptors, resulting in 1826 descriptors per molecule. Therefore, each ionic liquid, which includes both a cation and an anion, initially has 3652 descriptors.

To refine our feature set for machine learning applications, we implemented a feature reduction process. This involved removing descriptors that were consistently zero across all samples, as these add no discriminative information. Additionally, we excluded any features that contained NaN or infinite values to ensure the integrity of our dataset. After this cleanup process, the final number of features came to 1090 for cations and 672 for anions. Therefore, each ionic liquid in our study is effectively described by a total of 1762 features. 

Additionally, during the calculation of molecular descriptors, the Mordred package was unable to process the SMILES representations of 34 molecules. These instances were subsequently removed from our dataset. After excluding these molecules, the final dataset comprises 14 positive examples and 1748 negative examples.

\subsection{Algorithms}
In this subsection, we describe the algorithms employed to construct our classifier. Specifically, we utilized Logistic Regression (LR), Support Vector Machine (SVM), Random Forest (RF), and Extreme Gradient Boosting (XGBoost). The LR, SVM, and RF algorithms were implemented using the scikit learn package \cite{pedregosa2011scikit}. XGBoost was implemented using its own library \cite{chen2016xgboost}.

\subsubsection{Logistic Regression}
Logistic Regression (LR) is a statistical model used primarily for binary classification tasks. It estimates the probability that a given input belongs to a particular category by using a logistic function. The logistic regression classifier predicts whether an instance is positive (+1) or negative (-1). The logistic function transforms the output of a linear equation into a probability score, which ranges from 0 to 1. The probability $P$ that an input $x$ belongs to the default class (class 1) can be expressed mathematically as:
$$
P(y|x)=\frac{1}{1+e^{-y\cdot(w^T\cdot x +b)}}\,
$$
where $w$  is the weight vector representing the coefficients of the input features, and $b$ is the bias term. If the calculated probability is 0.5 or higher, the model predicts the instance as positive (+1); if it is below 0.5, it predicts the instance as negative (-1). This threshold-based decision rule allows LR to provide a clear binary classification based on the defined criteria. A diagram representing the LR classifier is shown in Fig.~\ref{fig:algorithms-sub1}.

In implementing the Logistic Regression model, we conducted a grid search to optimize several hyperparameters. This search involved varying the regularization strength to help prevent overfitting and testing two types of penalty norms: L1 and L2. Additionally, we evaluated two solvers: `liblinear' and `saga'. Through this grid search, we explored various combinations of regularization strengths, penalty types, and solver options, assessing their impact on the model’s accuracy and robustness

\subsubsection{Support Vector Machine}

Support Vector Machine (SVM) is a supervised learning model used for classification and regression. For classification tasks, such as ours, SVM attempts to maximize the margin $\gamma$,  defined as the distance between the nearest data points of each class (support vectors) and the hyperplane. To address datasets that are not linearly separable, SVM incorporates slack variables, which allow for some data points to violate the margin constraints. This flexibility enables SVM to fit more complex datasets by softening the margin, thus accommodating data points that lie within the margin or on the wrong side of the hyperplane. Additionally, SVM can be enhanced with the use of kernel functions, which allow the model to operate in a higher-dimensional space without explicitly computing the coordinates of the data in that space. Kernel functions transform the input data into a higher-dimensional space where a linear separator might be more effective. Common kernels include the linear, polynomial, radial basis function (RBF), and sigmoid. The utilization of kernels enables the SVM to capture complex relationships between the data points by increasing the dimensionality of the input space, thereby improving the classification performance even on data that is not linearly separable in their original dimensionality \cite{suthaharan2016support}. A diagram representing the SVM classifier is shown in Fig.~\ref{fig:algorithms-sub2}.

To optimize the SVM's performance, we conducted a grid search to fine-tune key hyperparameters. This included exploring various regularization strengths to determine the best trade-off between complexity and error, testing different kernel functions to see which best captured the data's underlying patterns, and varying the polynomial degree when using the polynomial kernel.

\subsubsection{Random Forest}
Random Forest (RF) is an ensemble learning method that combines multiple decision trees to improve classification accuracy and prevent overfitting. In Random Forest, each tree in the ensemble is built from a randomly selected subset of the training data and features, making each tree slightly different. When making predictions, the Random Forest classifier aggregates the decisions from all trees in the ensemble through a majority voting mechanism, which typically results in higher accuracy than any single decision tree could achieve \cite{biau2016random}.
A diagram representing the RF classifier is shown in Fig.~\ref{fig:algorithms-sub3}.
To enhance the performance of the Random Forest classifier, we conducted a grid search to optimize various hyperparameters. This process involved adjusting the number of trees in the forest, the maximum depth of each tree, the minimum number of samples required to split an internal node, and the minimum number of samples required at a leaf node.

\subsubsection{XGboost}

\begin{figure}[ht]
    \centering
    \begin{subfigure}[b]{0.45\textwidth}
        \centering
        \includegraphics[width=\textwidth]{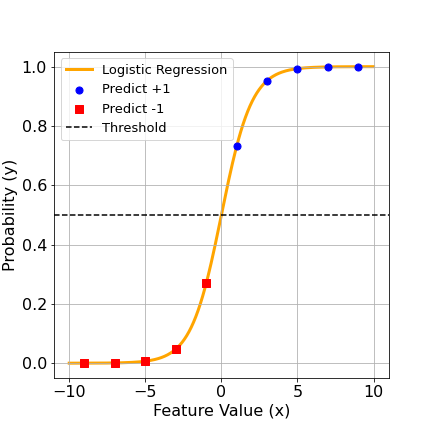}
        \caption{Logistic Regression.}
        \label{fig:algorithms-sub1}
    \end{subfigure}
    \hfill
    \begin{subfigure}[b]{0.45\textwidth}
        \centering
        \includegraphics[width=\textwidth]{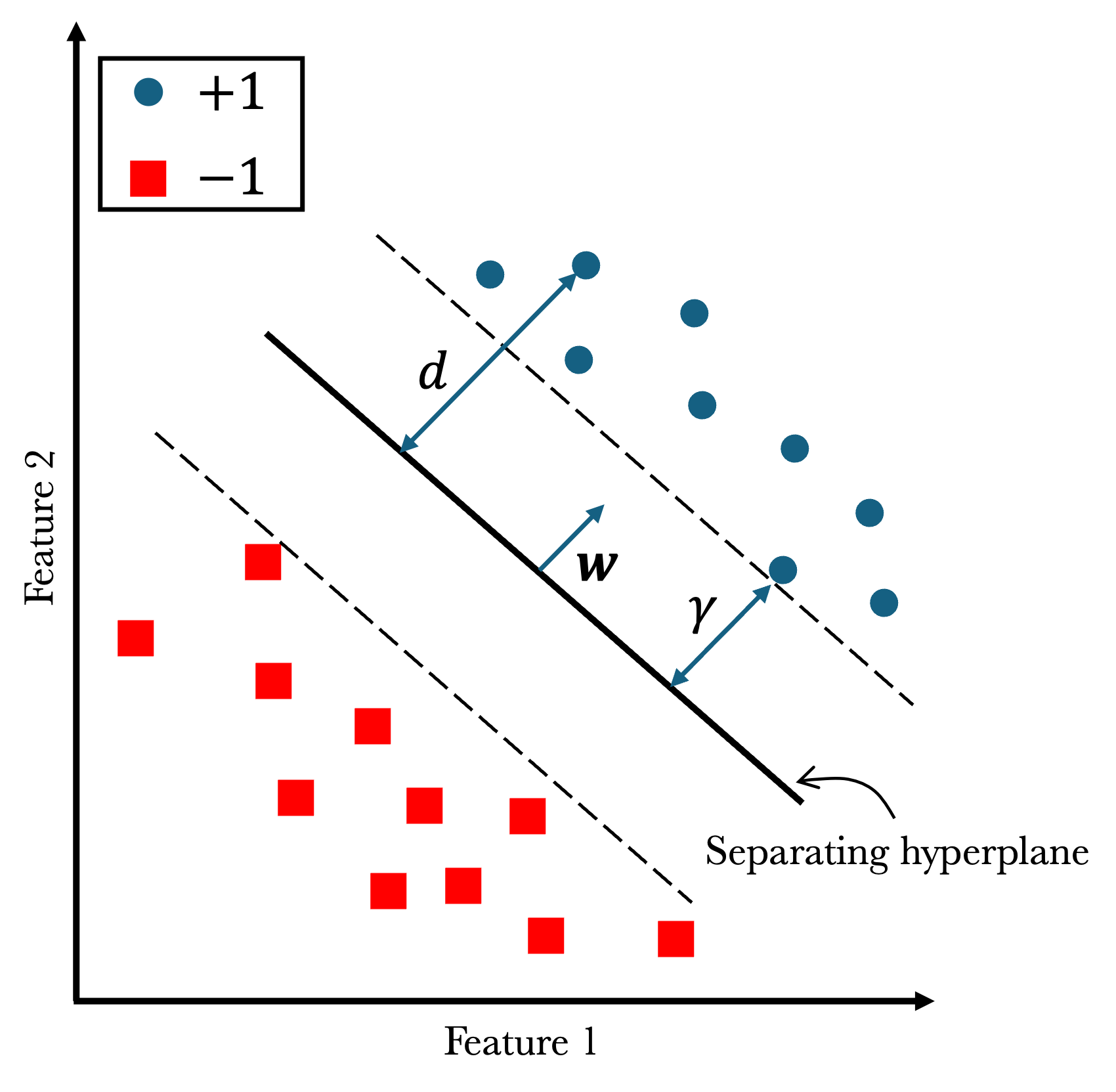}
        \caption{Support Vector Machine.}
        \label{fig:algorithms-sub2}
    \end{subfigure}
    \newline
    \begin{subfigure}[b]{0.45\textwidth}
        \centering
        \includegraphics[width=\textwidth]{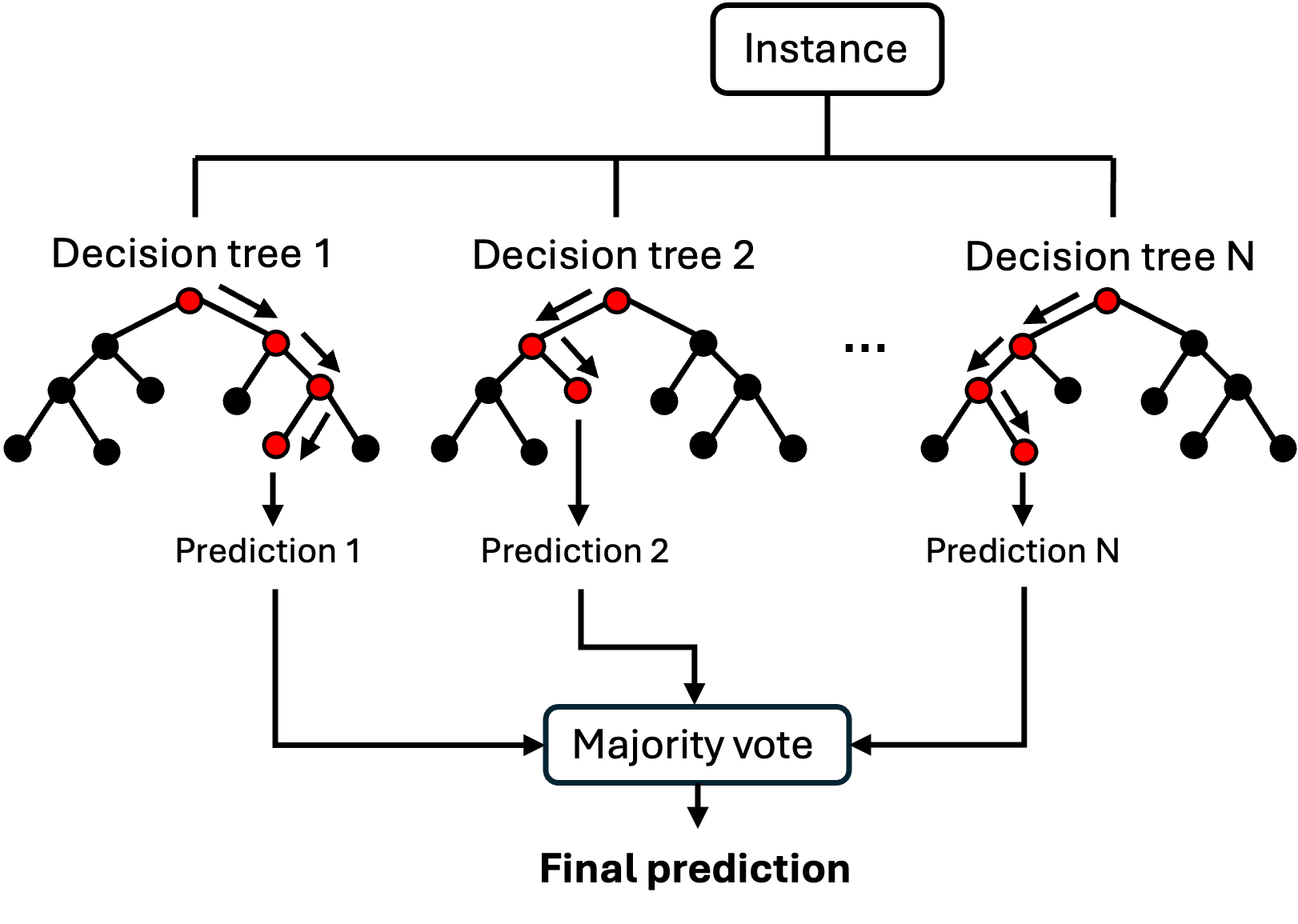}
        \caption{Random Forest.}
        \label{fig:algorithms-sub3}
    \end{subfigure}
    \hfill
    \begin{subfigure}[b]{0.49\textwidth}
        \centering
        \includegraphics[width=\textwidth]{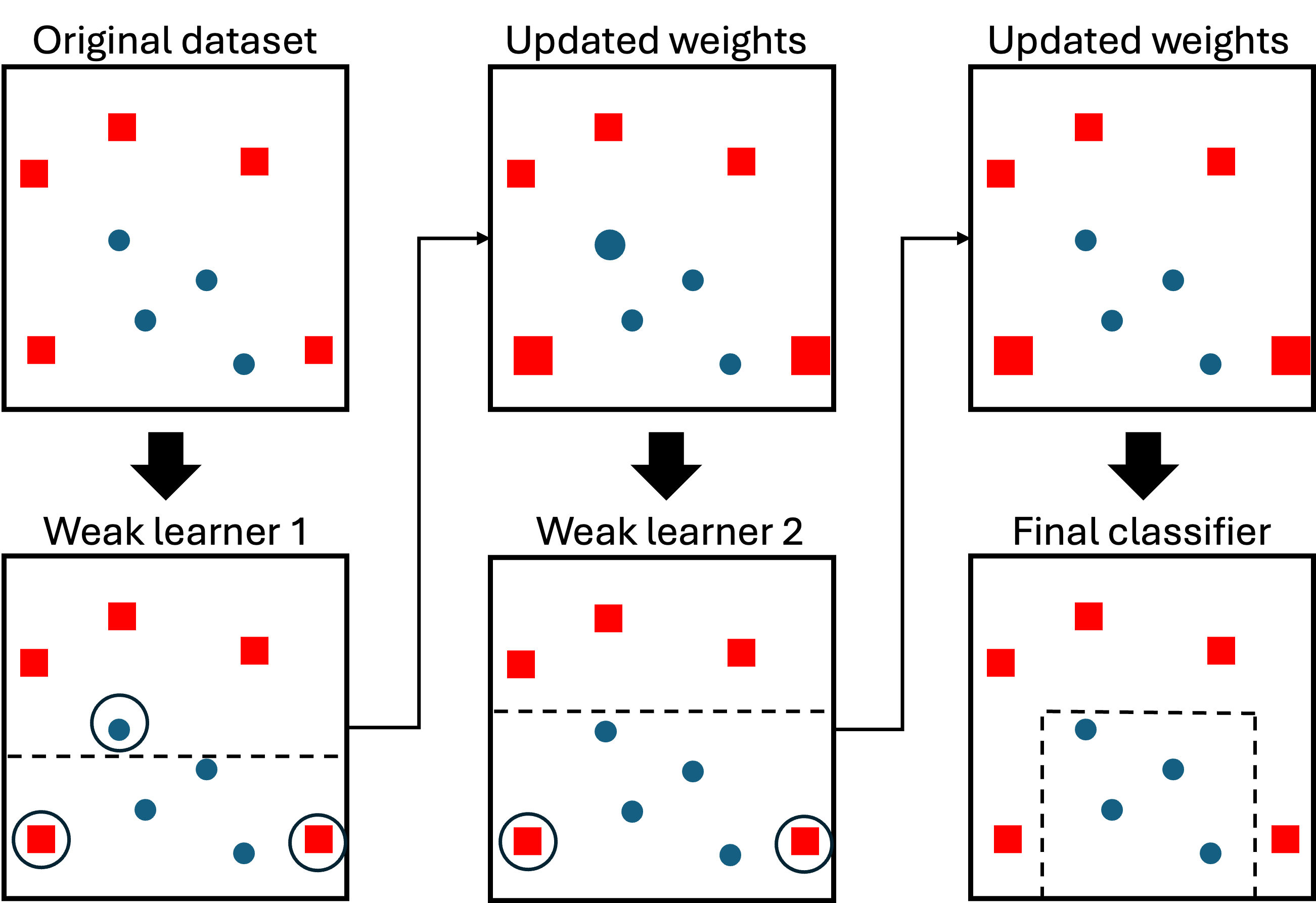}
        \caption{XGBoost.}
        \label{fig:algorithms-sub4}
    \end{subfigure}
    \caption{Diagrams representing (a) the logistic regression, (b) the support vector machine, (c) the random forest, and (d) the XGBoost classifiers.  }
    \label{fig:algorithms}
\end{figure}

XGBoost (eXtreme Gradient Boosting) is a sophisticated machine learning algorithm designed  for classification tasks. It operates by creating a series of simple models, known as weak learners (typically decision trees), which are sequentially improved to correct previous mistakes. Each new tree focuses on addressing errors from the earlier models, and when combined, these weak learners form a strong predictive model. The way this correction is implemented is by giving more weight to the training instances that were misclassified by the previous models. This method ensures that subsequent learners focus more on the difficult cases. As each new learner is added, it specifically addresses these weighted errors, refining the overall model's accuracy step by step \cite{chen2016xgboost}. A diagram representing the XGBoost classifier is shown in Fig.~\ref{fig:algorithms-sub4}.

To optimize the XGBoost model, we conducted a grid search that varied several key hyperparameters. This grid search method allowed us to systematically explore different configurations of the model, including adjustments to the number of trees, learning rate, tree depth, and minimum child weight. This approach helped us identify the most effective settings for our specific classification task.

\subsection{Performance Metrics}
To evaluate the performance of a classification model, several key metrics can be used: accuracy, precision, recall, and the F1 score.

\begin{itemize}
\renewcommand\labelitemi{} 
    \item \textbf{Accuracy} measures the proportion of true results (both true positives and true negatives) among the total number of cases examined. It is calculated as:
    \[
    \text{Accuracy} = \frac{\text{TP} + \text{TN}}{\text{TP} + \text{TN} + \text{FP} + \text{FN}}\,,
    \]
    where TP, TN, FP, and FN represent the numbers of true positives, true negatives, false positives, and false negatives, respectively. However, accuracy is not a reliable metric for datasets with imbalanced class distributions such as ours. Under such circumstances, a model may exhibit high accuracy simply by predominantly predicting the majority class, thereby failing to accurately reflect the true predictive performance with respect to the minority class.

    \item \textbf{Precision} measures the accuracy of positive predictions. It is defined as the ratio of true positives to the total predicted positives:
    \[
    \text{Precision} = \frac{\text{TP}}{\text{TP} + \text{FP}}\,.
    \]
    This metric highlights the reliability of the positive class identification.

    \item \textbf{Recall} indicates the ability of a model to find all the relevant cases (positive class) within a dataset. Mathematically, it is the ratio of true positives to the actual number of positives:
    \[
    \text{Recall} = \frac{\text{TP}}{\text{TP} + \text{FN}}\,.
    \]
    This metric focuses on the coverage of actual positive samples.

    \item \textbf{F1 Score} is the harmonic mean of precision and recall, providing a balance between them. It is particularly useful when the class distribution is uneven. The F1 score is given by:
    \[
    \text{F1 Score} = 2 \times \frac{\text{Precision} \times \text{Recall}}{\text{Precision} + \text{Recall}}\,.
    \]
    This score is a better measure when the data shows a significant imbalance between the classes.
\end{itemize}

For this study, we choose to use the F1 Score as the primary metric to assess the performance of our models. The F1 Score is selected because it balances the precision and recall of the model, providing a more holistic view of its effectiveness.

\subsection{Treatment of Imbalanced Data}
Our dataset exhibits a significant class imbalance with only 14 positive examples compared to 1748 negative examples, constituting approximately 0.79\% of the dataset as positive. Such imbalance can lead to biased model predictions that favor the majority class. To address this issue, we applied a combination of oversampling the minority class and undersampling the majority class.

\begin{figure}[ht!]
    \centering
    \includegraphics[width=0.45\linewidth]{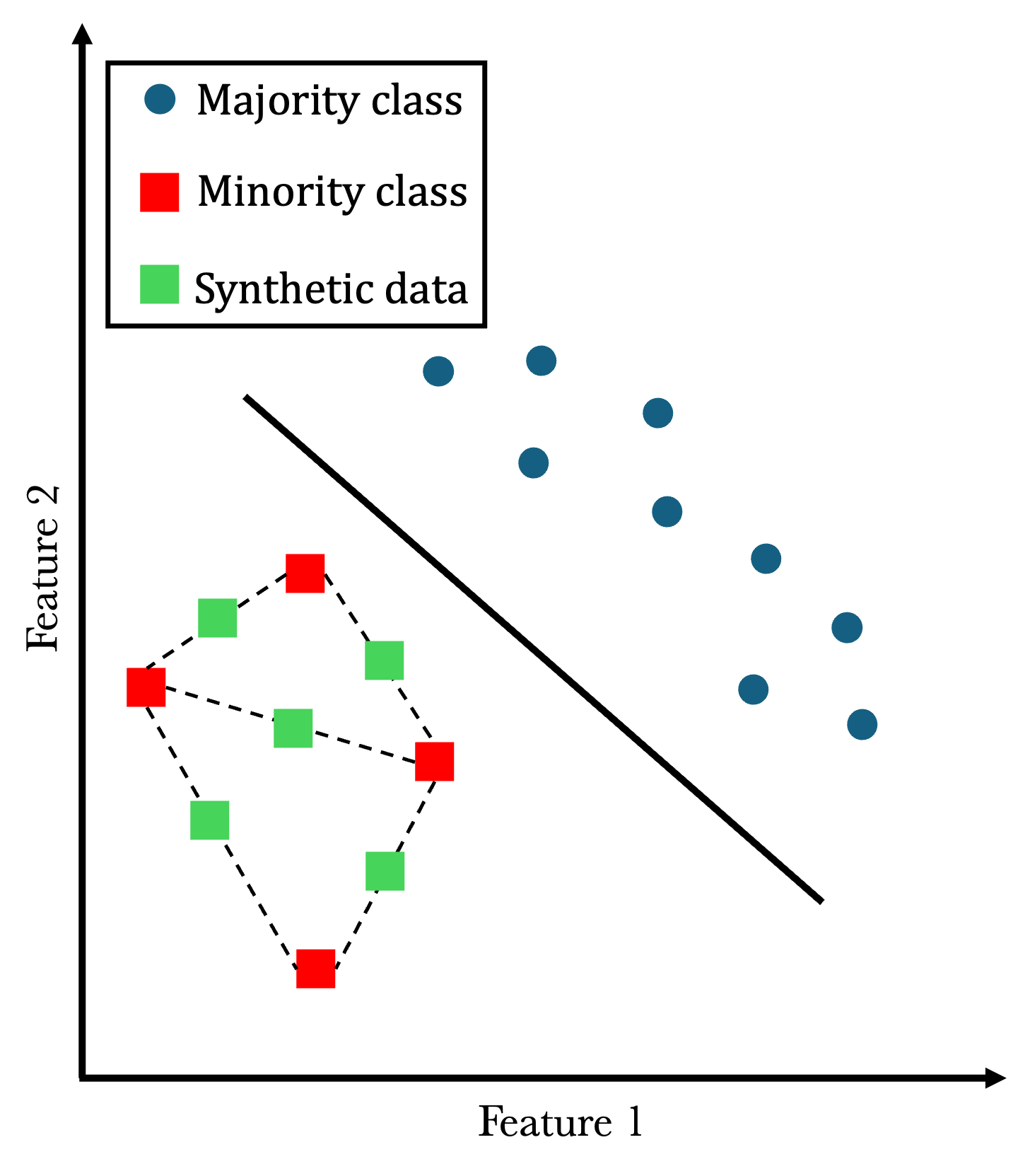}
    \caption{Visualization of the SMOTE algorithm using the 2 nearest neighbors of the minority class.}
    \label{fig:smote}
\end{figure}

Specifically, we employed the Synthetic Minority Over-sampling Technique (SMOTE) for oversampling. SMOTE works by creating synthetic samples from the minority class instead of creating copies. It selects two or more similar instances (using a distance measure) and perturbing an instance one feature at a time by a random amount within the difference to the neighboring instances. This approach not only increases the number of instances in the minority class but also introduces minor variations, making the overfitting less likely compared to simple replication \cite{chawla2002smote}. Figure \ref{fig:smote} illustrates the implementation of the SMOTE algorithm using the two nearest neighbors of each minority class instance to generate new synthetic data.

Following the application of SMOTE, we also implemented random undersampling of the majority class to balance the class distribution further. This combination helps in achieving a more balanced dataset, which enhances the training process and leads to a more generalized model that performs better on unseen data.

To optimize the resampling strategy, a grid search was conducted on the parameters associated with SMOTE. This grid search explored different configurations for the sampling strategies and the number of neighbors considered in the SMOTE algorithm. We varied the proportions in which the minority class should be oversampled and the majority class undersampled, alongside adjusting the number of nearest neighbors to use when generating synthetic samples.
After addressing the class imbalance through resampling techniques, we proceeded to scale the features of the dataset. Scaling is an essential preprocessing step that standardizes the range of the features, ensuring that no single feature dominates the model due to its larger scale.

\subsection{Cross-validation}

To evaluate our models, we divided the data into two sets: 80\% for training and 20\% for testing. The training set comprises 11 positive examples and 1398 negative examples, while the testing set contains 3 positive examples and 350 negative examples. Within the 80\% training subset, we further employed a 5-fold cross-validation method. Cross-validation is a statistical technique used to estimate the performance of machine learning models. It involves partitioning the training dataset into a number of folds. In 5-fold cross-validation, the dataset is split into five equally sized folds. In each iteration, four folds are used for training the model and the fifth fold is used as a validation set to assess model performance. This process is repeated five times, with each fold being used as the validation set once. Cross-validation helps in ensuring that our model's performance is reliable and not overly dependent on any particular subset of the training data, thus preventing overfitting and promoting a model that generalizes well to new unseen data. The hyperparameters explored during the grid search for each model, as well as those associated with the undersampling and oversampling strategies, are reported in the Appendix in Tables~\ref{tab:models_hyperparameters} and \ref{tab:smote_hyperparameters} respectively.

\section{Results} \label{sec:results}

This section presents the outcomes of our study, detailing the performance of the various machine learning models utilized, the discovery of new propellants from the ionic liquid (IL) database, and the application of our findings to IL-like molecules. We also discuss the importance of different features in the predictive models, providing insights into the factors that significantly influence the identification of suitable propellants.

\subsection{Performance of the algorithms}

Given the stochastic nature of the resampling methods, we conducted 50 iterations for each model, varying the random state of both SMOTE and the random undersampling in each iteration. For each iteration, we recorded the precision, recall, and F1 scores on both the training and testing sets. This robust evaluation allowed us to average these metrics across all iterations, providing a more reliable performance assessment. The results for each model, encompassing precision, recall, and F1 score are comprehensively detailed in Table~\ref{tab:performance_metrics}. The table provides a clear view of how each model performs on both the training and testing sets. The optimal parameters for each model, selected based on the F1 score, are shown in the Appendix in Table~\ref{tab:optimal_parameters}.

\begin{table}[H]
\centering
\caption{Performance metrics for Logistic Regression, Support Vector Machine, Random Forest, and XGBoost models.}
\begin{tabular}{ll|c|c|c|c|}
\cline{3-6}
                                                &                    & \textbf{LR} & \textbf{SVM} & \textbf{RF} & \textbf{XGBoost} \\ \hline
\multicolumn{1}{|l|}{\multirow{3}{*}{Training}} & \textbf{Precision} & 0.24        & 0.56         & 0.29        & 0.31             \\ \cline{2-6} 
\multicolumn{1}{|l|}{}                          & \textbf{Recall}    & 1.0         & 1.0          & 0.86        & 1.0              \\ \cline{2-6} 
\multicolumn{1}{|l|}{}                          & \textbf{F1}        & 0.39        & 0.72         & 0.43        & 0.48             \\ \hline
\multicolumn{1}{|l|}{\multirow{3}{*}{Testing}}  & \textbf{Precision} & 0.17        & 0.45         & 0.26        & 0.25             \\ \cline{2-6} 
\multicolumn{1}{|l|}{}                          & \textbf{Recall}    & 1.0         & 0.99         & 0.96        & 0.91             \\ \cline{2-6} 
\multicolumn{1}{|l|}{}                          & \textbf{F1}        & 0.29        & 0.61         & 0.41        & 0.39             \\ \hline
\end{tabular}
\label{tab:performance_metrics}
\end{table}

Among the evaluated models, the SVM demonstrated the highest precision and F1 score in both the training and testing phases. Given its superior performance, the SVM classifier has been selected for application on unknown ionic liquids within our dataset in order to discover new candidate propellants.

\subsection{Newly Discovered Propellants from IL Database}
In this section, we focus on the 465 ionic liquids with missing data on density, viscosity, or surface tension, which had prevented their initial classification. Molecular descriptors for these liquids were computed using the Mordred package. These descriptors were then used as inputs for our best-performing classifier, the SVM model, which was selected based on its superior F1 score. The classification results revealed that out of the 465 ionic liquids, 19 were identified as positive examples potentially suitable as propellants, which are reported in Table~\ref{tab:19pos}.

\begin{table}[ht!]
\centering
\caption{List of positively predicted ionic liquids.}
\renewcommand{\arraystretch}{1.2}
\begin{tabular}{|l|l|}
\hline
\textbf{Cation}                             & \textbf{Anion}                                                                                       \\ \hline
1,3-dimethyl-1H-imidazol-3-ium              & (pentafluoroethyl)trifluoroboranuide                                                                 \\ \hline
1,3-dimethyl-1H-imidazol-3-ium              & (trifluoromethyl)trifluoroboranuide                                                                  \\ \hline
1-(2-methoxyethyl)-1-methylpyrrolidin-1-ium & bis(flurosulfonyl)azanide                                                                            \\ \hline
1-(2-methoxyethyl)pyridin-1-ium             & bis(trifluoromethylsulfonyl)azanide                                                                  \\ \hline
1-butyl-3-methyl-1H-imidazol-3-ium          & tetrachlorogallate                                                                                   \\ \hline
1-ethyl-1H-imidazol-3-ium                   & \begin{tabular}[c]{@{}l@{}}(pentafluroethylsulfonyl)(fluorosulfonyl)\\ azanide\end{tabular}          \\ \hline
1-ethyl-3-methyl-1H-imidazol-3-ium          & \begin{tabular}[c]{@{}l@{}}(pentafluroethylsulfonyl)\\ (trifluoromethylsulfonyl)azanide\end{tabular} \\ \hline
1-ethyl-3-methyl-1H-imidazol-3-ium          & (trifluoromethyl)trifluoroboranuide                                                                  \\ \hline
1-ethyl-3-methyl-1H-imidazol-3-ium          & \begin{tabular}[c]{@{}l@{}}2,2,2-trifluoro-N-(trifluoromethylsulfonyl) \\ acetamide\end{tabular}     \\ \hline
1-ethyl-3-methyl-1H-imidazol-3-ium          & bis(flurosulfonyl)azanide                                                                            \\ \hline
1-ethyl-3-methyl-1H-imidazol-3-ium          & fluoranesulfonate                                                                                    \\ \hline
1-ethyl-3-methyl-1H-imidazol-3-ium          & tetrachloroferrate (III)                                                                             \\ \hline
1-ethyl-3-methyl-1H-imidazol-3-ium          & tetrachloroindate                                                                                    \\ \hline
1-ethyl-3-methyl-1H-imidazol-3-ium          & bis(trifluoromethylsulfonyl)azanide                                                                  \\ \hline
1-ethyl-3-methylpyridin-1-ium               & bis(trifluoromethylsulfonyl)azanide                                                                  \\ \hline
1-ethyl-4-methylpyrrolidin-1-ium            & bis(flurosulfonyl)azanide                                                                            \\ \hline
1-methyl-1-propylpyrrolidin-1-ium           & bis(trifluoromethylsulfonyl)azanide                                                                  \\ \hline
1-methyl-3-(propan-2-yl)-1H-imidazol-3-ium  & tetrachloroaluminate                                                                                 \\ \hline
1-methyl-3-propyl-1H-imidazol-3-ium         & tetrachloroaluminate                                                                                 \\ \hline
\end{tabular}

\label{tab:19pos}
\end{table}

\subsection{Application to constructed IL-like Molecules}

In our efforts to identify candidate propellants, we initially categorized 14 ionic liquids as suitable based on their physical properties and discovered an additional 19 through predictions made by our SVM classifier. To expand our search further, we constructed a larger dataset composed of pairs from the 942 unique cations and 268 unique anions found in the original dataset, resulting in a total of 252,456 unique IL-like molecules. The term `IL-like molecules' refers to pairs of cations and anions that are not guaranteed to be actual ionic liquids.

\begin{figure}[H]
    \centering
    \includegraphics[width=\linewidth]{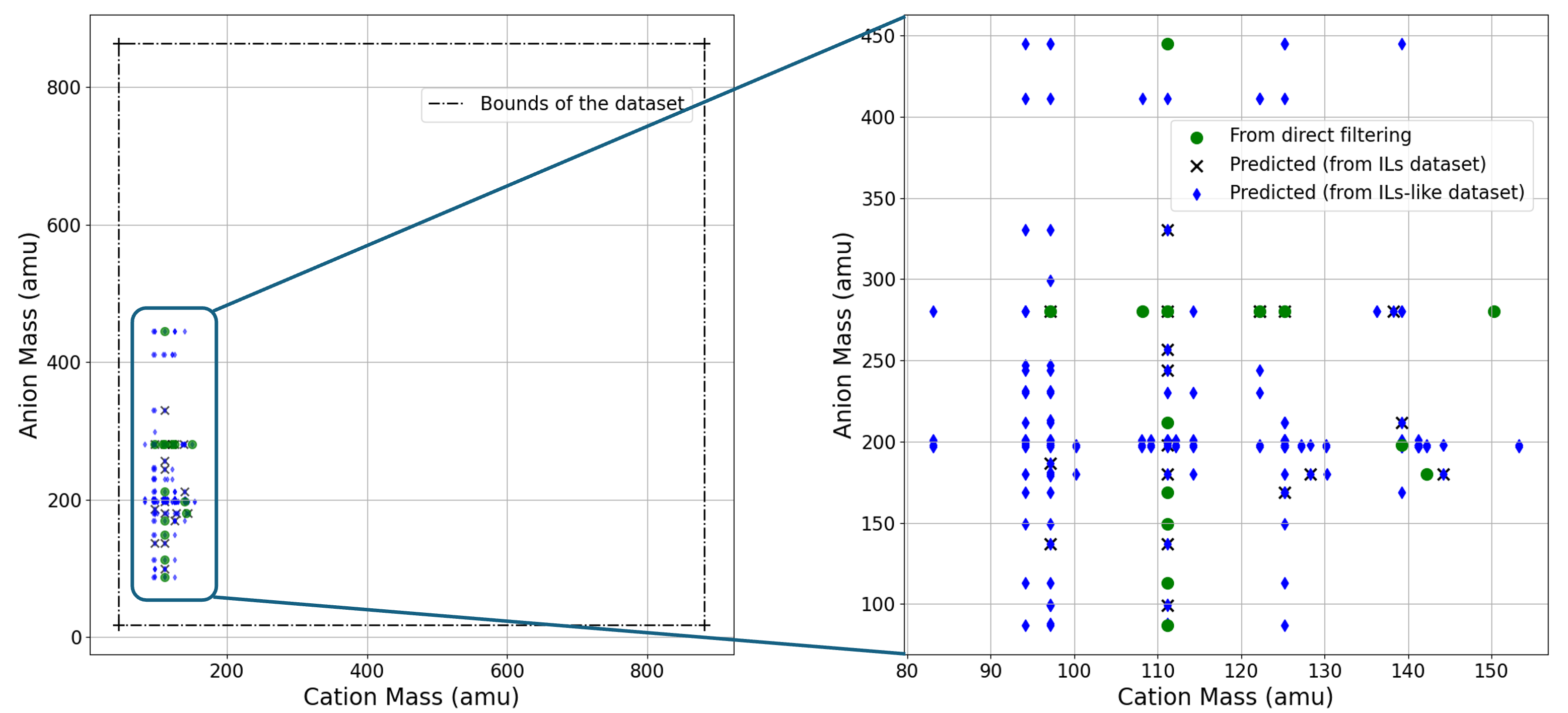}
    \caption{Positively predicted ionic liquids represented by their cation and anion masses.}
    \label{fig:discovered193}
\end{figure}

Molecular descriptors were calculated for each pair in this constructed dataset, after which the SVM classifier was applied to assess their suitability as propellants. From this analysis, we identified 193 positive examples. These can be categorized into three distinct groups: (1) the 14 initially identified through direct physical property filtering, (2) 19 discovered from predictions based on the original dataset, and (3) 160 discovered from the constructed dataset of IL-like molecules. The positively predicted ionic liquids are represented by the mass of their anions and cations in Figure~\ref{fig:discovered193}.

The 193 newly discovered candidate propellants are composed of pairs of 38 unique cations and 32 unique anions. The top 10 most frequently occurring cations and anions among these candidates are detailed in Table~\ref{tab:top10}.

\begin{table}[H]
    \centering
        
    \caption{Top 10 of the cations and the anions with the highest count in the set of positively predicted ionic liquids.}
\begin{tabular}{|>{\raggedright\arraybackslash}p{6cm}|c||>{\raggedright\arraybackslash}p{6.3cm}|c|}
\hline
\textbf{Cation Name} & \textbf{Count} & \textbf{Anion Name} & \textbf{Count} \\
\hline
1,3-dimethyl-1H-imidazol-3-ium & 31 & tetrachloroferrate (III) & 29 \\ \hline
1-ethyl-3-methyl-1H-imidazol-3-ium & 24 & tretrachloromanganese (II) & 25 \\ \hline
1-methylpyridin-1-ium & 21 & bis(trifluoromethylsulfonyl)azanide & 21 \\ \hline
1-methyl-3-(propan-2-yl)-1H-imidazol-3-ium & 12 & tetrachlorocobaltate (II) & 14 \\ \hline
1-methyl-3-propyl-1H-imidazol-3-ium & 9 & tetrachloronickelate (II) & 13 \\ \hline
1-butyl-3-methyl-1H-imidazol-3-ium & 8 & bis(flurosulfonyl)azanide & 10 \\ \hline
1,3-diethyl-1H-imidazol-3-ium & 8 & trifluorotris(pentafluoroethyl)-$\lambda^5$-phosphanuide & 8 \\ \hline
1-ethyl-1-methylpyrrolidin-1-ium & 7 & tris(trifluoromethylsulfonyl)methide & 7 \\ \hline
1-ethyl-4-methylpyridin-1-ium & 6 & tetrachloroaluminate & 7 \\ \hline
1,3-dimethylpyridin-1-ium & 5 & tetrachlorogallate & 7 \\
\hline
\end{tabular}

    \label{tab:top10}
\end{table}

\subsection{Features Importance}

To evaluate the contribution of individual features to the model's predictions, we employed SHapley Additive exPlanations (SHAP) values, a method from cooperative game theory. SHAP values provide a measure of the importance of each feature in the prediction of a model for every single observation, rather than providing a general metric of feature importance across the entire dataset. These values are particularly useful for understanding the decision-making process of complex models by attributing the prediction output to each input feature \cite{lundberg2017unified}.

In our analysis, we calculated SHAP values for a subset of our training dataset, comprising 50 examples—39 negative and 11 positive cases. Given the extensive number of features in our dataset, calculating SHAP values is computationally expensive. Limiting our analysis to a smaller subset allows for a manageable computation while still providing meaningful insights. The resultant SHAP values allowed us to identify and rank the features according to their influence on the classifier’s output, offering insights into which properties of ionic liquids most significantly impact their suitability as propellants.

\begin{figure}[ht!]
    \centering
    \includegraphics[width=0.8\linewidth]{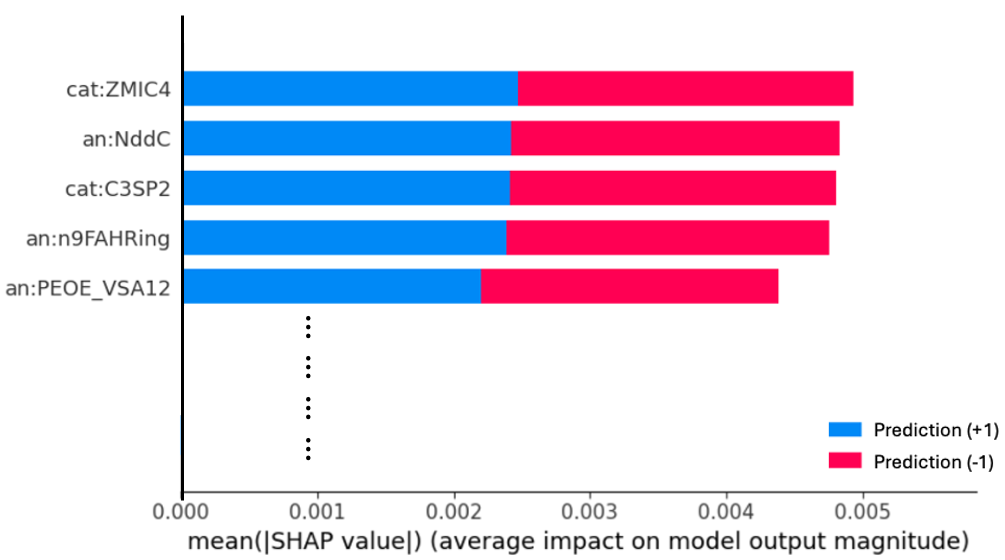}
    \caption{SHAP values of the top 5 most influential molecular descriptors.}
    \label{fig:shap}
\end{figure}

Figure~\ref{fig:shap} presents a summary plot of SHAP values for the top 5 most influential features that impact the SVM classifier's predictions on a subset of our dataset. This figure illustrates the relative contribution of each feature towards the model's output. Notably, features prefixed with `cat:' denote properties of cations, while those beginning with `an:' represent properties of anions. For example, `cat:ZMIC4' represents the 4-ordered Z-modified information content in the cation. The Z-modified Information Content (ZMIC) is an information theory-based metric employed in molecular informatics to quantify the complexity of molecular structures in terms of information content, integrating both topological characteristics of chemical graphs and stereochemical attributes \cite{king1989az}. In addition, `an:NddC' represents the number of double bonds connected to carbon atoms in the anion. Further, `cat:C3SP2' indicates the count of sp2 hybridized carbon atoms in the cation that are bonded to three other carbon atoms. The descriptor `an:n9FAHRing' measures the count of 9-membered aliphatic fused hetero rings in the anion. Additionally, `an:PEOE$\_$VSA12' is used to measure the van der Waals surface area of atoms in anions with partial charges between 0.20 and 0.25.

\begin{figure}[H]
    \centering
    \includegraphics[width=\linewidth]{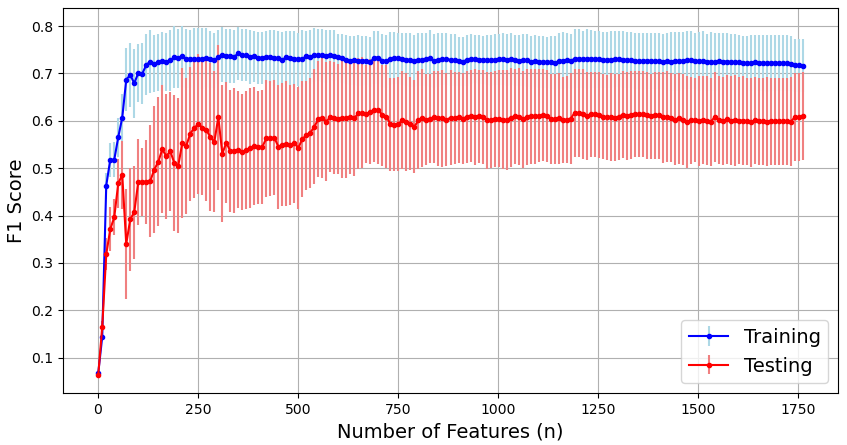}
    \caption{F1 score of training as a function of the number of features ranked by their SHAP values.}
    \label{fig:f1score}
\end{figure}

In order to assess the impact of feature selection on classifier performance, we first optimized the hyperparameters of the SVM classifier using a grid search based on the full set of 1762 features. With these optimized hyperparameters fixed, we then utilized the ranking provided by SHAP values to progressively retrain the classifier, starting with the single most influential feature and incrementally adding features. For each configuration, the classifier was retrained, and the F1 scores for the training set were recorded. When incrementing the number of features from 1 to 1762, at each step, we repeated the evaluation 50 times with different seed numbers for SMOTE and the random undersampling. This iterative process is illustrated in Fig.~\ref{fig:f1score}, which depicts the average F1 score along with the error bars corresponding to one standard deviation, as a function of the number of features used. Analysis of this figure indicates that the F1 score of testing converges to approximately 0.60 when utilizing the top 700 features. This observation suggests that beyond the first 700 features, additional features contribute minimally to enhancing the model's predictive accuracy, highlighting the potential for substantial model simplification without significant loss of performance.

\section{Discussion} \label{sec:discussion}

One significant outcome of this study is the identification of 193 candidate ILs, expanding the pool of potential propellants. This underscores the utility of molecular descriptors and machine learning in extrapolating the properties of ILs with incomplete experimental data. However, given the testing precision of our SVM model at 45\%, we estimate that among the 19 candidates discovered from predictions based on the original dataset, approximately 9 are likely to be true positives, whereas the remaining 10 could be false positives. Similarly, among the 160 candidates discovered from the constructed dataset, 72 are likely to be true positives, whereas the remaining 88 could be false positives. Adding these 81 true positives (9 from the original dataset predictions and 72 from the constructed dataset predictions) to the 14 candidates identified through direct physical property filtering, we estimate the total number of potentially true positive candidates to be 95.

In our constructed dataset of IL-like molecules, the masses of cations vary from 46.1 to 881.6 atomic mass units (amu), while the masses of anions range from 17.01 to 863.21 amu. However, for the positively predicted ionic liquids, we observed a narrower mass range: cations between 83.11 and 153.24 amu, and anions between 86.81 and 445.01 amu. This narrower range in positively predicted examples is attributed to the restricted mass range of the training set used to develop the SVM classifier. The classifier's predictions are primarily reliable within this trained mass range, as the SVM model is limited in its ability to extrapolate beyond the training data. Consequently, there may be actual suitable propellants with mass characteristics far outside this range that our classifier could not effectively evaluate. 

The analysis of features importance using SHAP values indicated that the top 700 physical descriptors are sufficient to reconstruct the classifier. Among these 700 most important features, 433 are cation features and 267 are anion features. This distribution suggests that the properties of cations are slightly more influential than those of anions in determining the classification outcomes.

The study's methodology and findings have broader implications beyond the specific application of electrospray propulsion. The machine learning framework developed here can be adapted to other fields requiring the identification and characterization of compounds with specific physical properties, such as pharmaceuticals, materials science, and chemical engineering. By leveraging computational techniques to predict molecular properties, we can significantly accelerate the discovery and optimization process.

\section{Future Work} \label{sec:future}
\begin{figure}[ht!]
    \centering
    \includegraphics[width=1\linewidth]{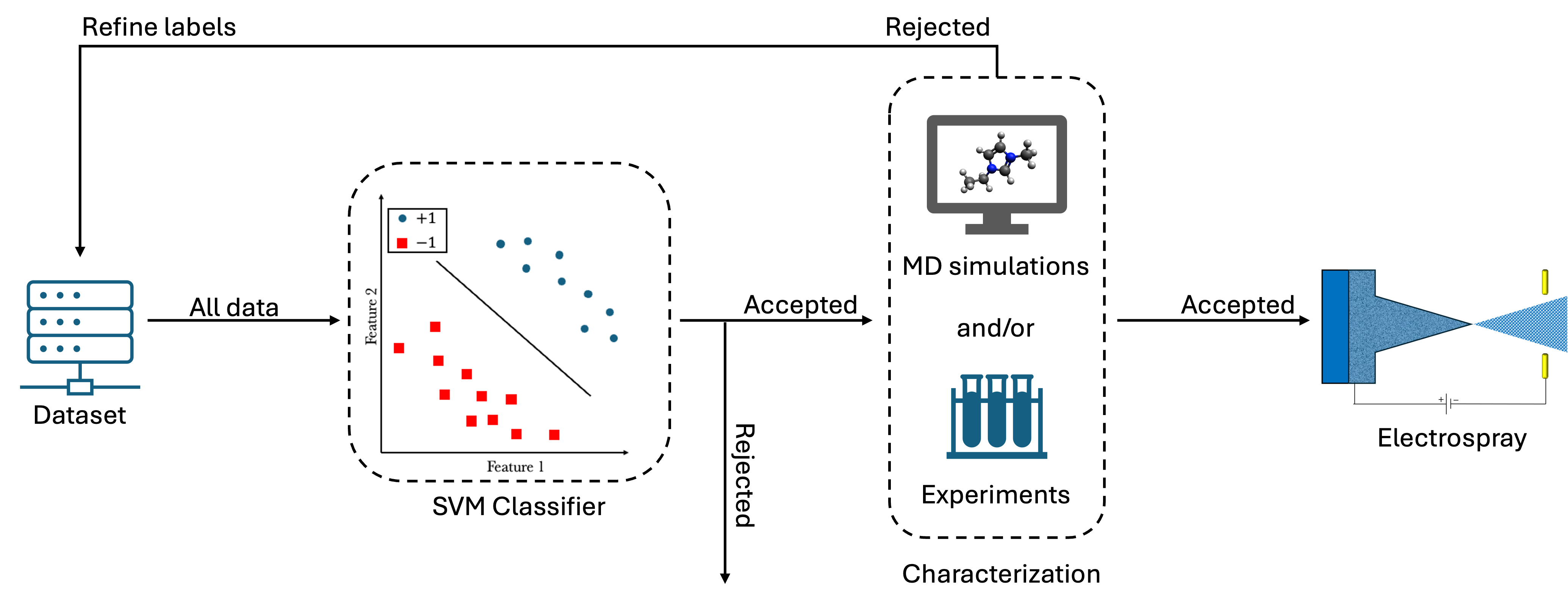}
    \caption{Diagram of the validation procedure of the predictions using MD simulations and experiments.}
    \label{fig:future}
\end{figure}

In future work, we aim to validate the machine learning predictions for ionic liquids by integrating Molecular Dynamics (MD) simulations and experimental approaches. The predictive accuracy of our SVM classifier will be tested against the outcomes of MD simulations. Concurrently, experimental validation will be conducted to confirm the feasibility and performance of these ionic liquids as propellants in practical applications. The results from both MD simulations and experiments will not only serve to verify our machine learning predictions but also contribute to refining the labeling of our dataset. This will establish a closed iterative loop, where empirical data continuously informs and enhances the accuracy of our predictive models. Such an approach will strengthen the reliability of our findings and potentially uncover new avenues for optimizing the properties of ionic liquids for specific applications. A diagram showing the proposed validation procedure is shown in Fig.~\ref{fig:future}.

In addition, several limitations need to be addressed. The exclusion of key properties such as electrical conductivity, Gibbs free energy, and electrochemical window from the dataset limits the comprehensiveness of the model's predictions. Future work should aim to incorporate these properties to provide a more holistic assessment of IL suitability.

\section{Conclusions} \label{sec:conclusion}
This study has effectively demonstrated the application of machine learning techniques to predict the suitability of ionic liquids as propellants for electrospray thrusters, which are critical in the micropropulsion systems of small satellites. Our approach utilized molecular descriptors derived from the SMILES representations of ionic liquids to train multiple classifiers: Logistic Regression, Support Vector Machine (SVM), Random Forest, and XGBoost, among which the SVM model exhibited superior performance based on F1 scores. Our machine learning framework involved rigorous data preprocessing, which included handling imbalanced datasets through strategic resampling techniques. Specifically, we applied oversampling of the minority class using the Synthetic Minority Over-sampling Technique (SMOTE) and undersampling of the majority class to achieve a balanced training dataset.

Through this machine learning framework, we successfully classified ionic liquids based on their physicochemical properties, highlighting 14 initially identified ionic liquids from the complete data set and discovering 19 additional candidates from the incomplete data set. Further explorations into constructed IL-like molecules have expanded our candidate list to 193 potential propellants. Moreover, the integration of SHAP values has provided deep insights into the decision-making processes of our models, identifying key molecular descriptors that influence the predictions. 

As we continue to refine our models and predictions, the next steps involve integrating Molecular Dynamics simulations and experimental validations to test the feasibility and performance of these ionic liquids as propellants. This collaborative approach between computational predictions and empirical validations will enhance the accuracy of our findings and may unveil new opportunities for optimizing ionic liquid properties for specific applications.

The methodology and insights derived from this study are not confined solely to electrospray propulsion or ionic liquids but are broadly applicable to various other fields requiring the identification and characterization of molecules with specific physical properties. By utilizing machine learning techniques to analyze molecular descriptors, this approach can be adapted to predict the suitability of compounds across a wide range of applications, from pharmaceuticals to energy storage materials.

\section*{Acknowledgements}
We thank the AFOSR Young Investigator Program FA9550-23-1-0141 under Dr. Justin Koo for supporting this work.

\appendix

\section{Grid search results} \label{sec:appendix_gridsearch}
\begin{table}[H]
\centering
\caption{Hyperparameters used in grid search for each model.}
\begin{tabular}{|l|l|}
\hline
\textbf{Model}             & \textbf{Parameters}                                                       \\ \hline
Logistic Regression        & \begin{tabular}[c]{@{}l@{}}logreg\_\_C: [0.01, 0.1, 1, 10, 100],\\ logreg\_\_penalty: [l1, l2],\\ logreg\_\_solver: [liblinear, saga]\end{tabular} \\ \hline
Support Vector Machine (SVM) & \begin{tabular}[c]{@{}l@{}}svm\_\_C: [0.1, 1, 10, 20],\\ svm\_\_kernel: [linear, rbf, poly],\\ svm\_\_degree: f\end{tabular} \\ \hline
Random Forest (RF)         & \begin{tabular}[c]{@{}l@{}}rf\_\_n\_estimators: [10, 50, 100],\\ rf\_\_max\_depth: [None, 10, 20, 30],\\ rf\_\_min\_samples\_split: [2, 5, 10],\\ rf\_\_min\_samples\_leaf: [1, 2, 4]\end{tabular} \\ \hline
XGBoost                    & \begin{tabular}[c]{@{}l@{}}xgb\_\_n\_estimators: [100, 200, 300],\\ xgb\_\_learning\_rate: [0.01, 0.1, 0.2],\\ xgb\_\_max\_depth: [3, 6, 9],\\ xgb\_\_min\_child\_weight: [1, 2, 3]\end{tabular} \\ \hline
\end{tabular}
\label{tab:models_hyperparameters}
\end{table}

\begin{table}[H]
\centering
\caption{Hyperparameters used in grid search for resampling.}
\begin{tabular}{|l|l|}
\hline
\textbf{Resampling Component}             & \textbf{Parameters}                                                       \\ \hline
Oversampling Strategy & smote\_over\_\_sampling\_strategy: [0.1, 0.5, 1] \\ \hline
Undersampling Strategy         & smote\_under\_\_sampling\_strategy: [0.1, 0.5, 1] \\ \hline
Number of Neighbors        & smote\_over\_\_k\_neighbors: [3, 4, 5] \\ \hline
\end{tabular}
\label{tab:smote_hyperparameters}
\end{table}

\begin{table}[H]
\centering
\caption{Optimal parameters for each model from the grid search.}
\begin{tabular}{|l|l|}
\hline
\textbf{Model} & \textbf{Optimal Parameters} \\ \hline
Logistic Regression & \begin{tabular}[c]{@{}l@{}}smote\_over\_\_k\_neighbors: 3, \\ smote\_over\_\_sampling\_strategy: 1, \\ smote\_under\_\_sampling\_strategy: 1, \\ logreg\_\_C: 0.01, \\ logreg\_\_penalty: l2, \\ logreg\_\_solver: liblinear\end{tabular} \\ \hline
SVM & \begin{tabular}[c]{@{}l@{}}smote\_over\_\_k\_neighbors: 4, \\ smote\_over\_\_sampling\_strategy: 0.5, \\ smote\_under\_\_sampling\_strategy: 1, \\ svm\_\_C: 20, \\ svm\_\_degree: 2, \\ svm\_\_kernel: rbf\end{tabular} \\ \hline
Random Forest & \begin{tabular}[c]{@{}l@{}}smote\_over\_\_k\_neighbors: 3, \\ smote\_over\_\_sampling\_strategy: 0.1, \\ smote\_under\_\_sampling\_strategy: 0.5, \\ rf\_\_max\_depth: 30, \\ rf\_\_min\_samples\_leaf: 4, \\ rf\_\_min\_samples\_split: 10, \\ rf\_\_n\_estimators: 50\end{tabular} \\ \hline
XGBoost & \begin{tabular}[c]{@{}l@{}}smote\_over\_\_k\_neighbors: 3, \\ smote\_over\_\_sampling\_strategy: 0.1, \\ smote\_under\_\_sampling\_strategy: 0.5, \\ xgb\_\_learning\_rate: 0.1, \\ xgb\_\_max\_depth: 6, \\ xgb\_\_min\_child\_weight: 1, \\ xgb\_\_n\_estimators: 300\end{tabular} \\ \hline
\end{tabular}
\label{tab:optimal_parameters}
\end{table}

 \label{sec:appendix_list}


\begin{thebibliography}{10}
\expandafter\ifx\csname url\endcsname\relax
  \def\url#1{\texttt{#1}}\fi
\expandafter\ifx\csname urlprefix\endcsname\relax\def\urlprefix{URL }\fi
\expandafter\ifx\csname href\endcsname\relax
  \def\href#1#2{#2} \def\path#1{#1}\fi

\bibitem{freemantle2010introduction}
M.~Freemantle, An introduction to ionic liquids, Royal Society of chemistry,
  2010.

\bibitem{greer2020industrial}
A.~J. Greer, J.~Jacquemin, C.~Hardacre, Industrial applications of ionic
  liquids, Molecules 25~(21) (2020) 5207.

\bibitem{welton2018ionic}
T.~Welton, Ionic liquids: a brief history, Biophysical reviews 10~(3) (2018)
  691--706.

\bibitem{hallett2011room}
J.~P. Hallett, T.~Welton, Room-temperature ionic liquids: solvents for
  synthesis and catalysis. 2, Chemical reviews 111~(5) (2011) 3508--3576.

\bibitem{zhou2009ionic}
F.~Zhou, Y.~Liang, W.~Liu, Ionic liquid lubricants: designed chemistry for
  engineering applications, Chemical Society Reviews 38~(9) (2009) 2590--2599.

\bibitem{shamshina2013ionic}
J.~L. Shamshina, P.~S. Barber, R.~D. Rogers, Ionic liquids in drug delivery,
  Expert opinion on drug delivery 10~(10) (2013) 1367--1381.

\bibitem{koutsoukos2021review}
S.~Koutsoukos, F.~Philippi, F.~Malaret, T.~Welton, A review on machine learning
  algorithms for the ionic liquid chemical space, Chemical science 12~(20)
  (2021) 6820--6843.

\bibitem{paduszynski2019extensive}
K.~Paduszynski, Extensive databases and group contribution qsprs of ionic
  liquids properties. 1. density, Industrial \& Engineering Chemistry Research
  58~(13) (2019) 5322--5338.

\bibitem{paduszynski2019extensive2}
K.~Paduszynski, Extensive databases and group contribution qsprs of ionic
  liquids properties. 2. viscosity, Industrial \& Engineering Chemistry
  Research 58~(36) (2019) 17049--17066.

\bibitem{paduszynski2021extensive}
K.~Paduszynski, Extensive databases and group contribution qsprs of ionic
  liquid properties. 3: surface tension, Industrial \& Engineering Chemistry
  Research 60~(15) (2021) 5705--5720.

\bibitem{constantinou1994new}
L.~Constantinou, R.~Gani, New group contribution method for estimating
  properties of pure compounds, AIChE Journal 40~(10) (1994) 1697--1710.

\bibitem{weininger1988smiles}
D.~Weininger, Smiles, a chemical language and information system. 1.
  introduction to methodology and encoding rules, Journal of chemical
  information and computer sciences 28~(1) (1988) 31--36.

\bibitem{pinheiro2020machine}
G.~A. Pinheiro, J.~Mucelini, M.~D. Soares, R.~C. Prati, J.~L. Da~Silva, M.~G.
  Quiles, Machine learning prediction of nine molecular properties based on the
  smiles representation of the qm9 quantum-chemistry dataset, The Journal of
  Physical Chemistry A 124~(47) (2020) 9854--9866.

\bibitem{chen2021predicting}
G.~Chen, L.~Tao, Y.~Li, Predicting polymers’ glass transition temperature by
  a chemical language processing model, Polymers 13~(11) (2021) 1898.

\bibitem{saini2023machine}
V.~Saini, Machine learning prediction of empirical polarity using smiles
  encoding of organic solvents, Molecular Diversity 27~(5) (2023) 2331--2343.

\bibitem{Krejci2015design}
D.~Krejci, F.~Mier-Hicks, C.~Fucetola, P.~C. Lozano, A.~H. Schouten, F.~Martel,
  \href{http://electricrocket.org/IEPC/IEPC-2015-149\_ISTS-2015-b-149.pdf}{{Design
  and Characterization of a Scalable ion Electrospray Propulsion System}},
  Joint Conference of 30th ISTS, 34th IEPC and 6th NSAT, Hyogo-Kobe, Japan
  (2015) 1--11.
\newline\urlprefix\url{http://electricrocket.org/IEPC/IEPC-2015-149\_ISTS-2015-b-149.pdf}

\bibitem{lozano2005ionic}
P.~Lozano, M.~Martinez-Sanchez, Ionic liquid ion sources: characterization of
  externally wetted emitters, Journal of colloid and interface science 282~(2)
  (2005) 415--421.

\bibitem{miller2019characterization}
C.~E. Miller, Characterization of ion cluster fragmentation in ionic liquid ion
  sources, Ph.D. thesis, Massachusetts Institute of Technology (2019).

\bibitem{moriwaki2018mordred}
H.~Moriwaki, Y.-S. Tian, N.~Kawashita, T.~Takagi, Mordred: a molecular
  descriptor calculator, Journal of cheminformatics 10 (2018) 1--14.

\bibitem{gallud2023studies}
X.~Gallud~Cidoncha, Studies on the physical structure, properties and operation
  of ionic liquid electrosprays in the pure-ion mode, Ph.D. thesis,
  Massachusetts Institute of Technology (2023).

\bibitem{landrum2006rdkit}
G.~Landrum, et~al., Rdkit: Open-source cheminformatics (2006).

\bibitem{pedregosa2011scikit}
F.~Pedregosa, G.~Varoquaux, A.~Gramfort, V.~Michel, B.~Thirion, O.~Grisel,
  M.~Blondel, P.~Prettenhofer, R.~Weiss, V.~Dubourg, et~al., Scikit-learn:
  Machine learning in python, the Journal of machine Learning research 12
  (2011) 2825--2830.

\bibitem{chen2016xgboost}
T.~Chen, C.~Guestrin, Xgboost: A scalable tree boosting system, in: Proceedings
  of the 22nd acm sigkdd international conference on knowledge discovery and
  data mining, 2016, pp. 785--794.

\bibitem{suthaharan2016support}
S.~Suthaharan, S.~Suthaharan, Support vector machine, Machine learning models
  and algorithms for big data classification: thinking with examples for
  effective learning (2016) 207--235.

\bibitem{biau2016random}
G.~Biau, E.~Scornet, A random forest guided tour, Test 25 (2016) 197--227.

\bibitem{chawla2002smote}
N.~V. Chawla, K.~W. Bowyer, L.~O. Hall, W.~P. Kegelmeyer, Smote: synthetic
  minority over-sampling technique, Journal of artificial intelligence research
  16 (2002) 321--357.

\bibitem{lundberg2017unified}
S.~M. Lundberg, S.-I. Lee, A unified approach to interpreting model
  predictions, Advances in neural information processing systems 30 (2017).

\bibitem{king1989az}
J.~W. King, Az-weighted information content index, International Journal of
  Quantum Chemistry 36~(S16) (1989) 165--170.

\end{thebibliography}





\end{document}